\documentclass[twocolumn,prl]{revtex4}
\usepackage{graphicx}
\usepackage{amsmath}
\usepackage[usenames]{color} 

\begin{document}
\title{Probing the Gravitational Wave Signature from Cosmic Phase Transitions at Different Scales}
\author{Lawrence M. Krauss$^1$}
\author{Katherine Jones-Smith$^2$}
\author{Harsh Mathur$^2$}
\author{James Dent$^1$}

\affiliation{$^1$School of Earth and Space Exploration and Physics Department, Arizona State University, Tempe AZ, 85287-1404}

\affiliation{$^2$CERCA, Department of Physics, Case Western Reserve University,10900 Euclid Avenue, Cleveland OH 44106-7079}

\begin{abstract}

We present a new signature by which to one could potentially discriminate between a spectrum of gravitational radiation generated by a self-ordering scalar field vs that of inflation, specifically a comparison of the magnitude of a flat spectrum at frequencies probed by future direct detection experiments to the magnitude of a possible polarization signal in the Cosmic Microwave Background (CMB) radiation.  In the process we clarify several issues related to the proper calculation of such modes, focusing on the effect of post-horizon-crossing evolution.

\end{abstract}

\maketitle

It was pointed out over 15 years ago that a continuous symmetry breaking phase transition, 
 can produce a scale invariant spectrum
of gravitational radiation similar to that of Inflation  \cite{lmk1992}. However, until prospects arose, via measurements of polarization of the CMB, for an actual measurement of this background, no detailed quantitative estimate of this effect had been performed. More recently a quantitative
calculation was carried out in the case of a scalar field that breaks a global O(N) symmetry \cite{prl2008}, confirming qualitative features and also
 demonstrating that the phase transition signal could be significantly larger than previously expected . 

Since the detection of primordial gravitational radiation is widely considered a smoking gun
for inflation it is important to carefully explore how to separate and disentangle this potentially 
competitive signal.  Baumann and Zaldarriaga \cite{baumann}
have recently confirmed that that the problem is more subtle than
one might initially suppose. The standard representation of CMB polarization data in terms of
spin-weighted harmonics of the $B$-mode would not be able to discriminate gravitational radiation produced by Inflation from that produced by a phase transition. Instead, in order to probe for superhorizon-sized correlations, they introduced
a new kind of $B$-mode dubbed the $\tilde{B}$ mode in ref \cite{baumann} and have argued that angular correlations on scales slightly larger than 2 degrees could distinguish an inflationary signature from that due to other phase transitions .  Subsequently Adshead and Lim \cite{lim,starobinsky} have pointed out that 
non-Gaussianity may provide another diagnostic. Both of these signatures will be challenging to extract from CMB data, however.

The purpose of this short note is is to demonstrate an in-principle clear and direct distinction 
between gravitational waves generated as a result of Inflation vs self-ordering, which derives from the fact that a scalar field associated with symmetry breaking continues to act as a source of gravitational waves inside the horizon in the latter case.  Primordial gravitational waves produce a CMB polarization signal due to anisotropic scattering of radiation by electrons in a quadrupole field, and this effect is dominated by modes which enter the horizon at decoupling, and again during reionization, both well into the matter-dominated era.  However, all currently proposed and envisaged direct interferometric gravitational wave detectors such as LISA and its projected successors will be sensitive to modes that entered the horizon in the radiation dominated era.  As we shall demonstrate, there will be  differing relative signatures for CMB signals vs those that might be observed in LISA or its successors in each of the two different cases (inflationary produced waves and those resulting from self-ordering scalar fields).

In order to calculate this difference one must carefully follow the dynamics of the scalar field inside the horizon.  In so doing we demonstrate the need to calculate effects well inside horizon crossing .  We also clarify confusion in the literature regarding the normalization conventions appropriate when comparing a CMB polarization signal and a direct sub-horizon gravitational strain signal, and  discuss the effects of vacuum energy domination on the evolution of the signals.

Like all forms of radiation, gravitational waves begin to oscillate as their period approaches the age of the universe and they come inside the horizon.  At the same time, they begin to redshift.   For gravitational waves from inflation, which are only sourced during the inflationary epoch, these are the only effects that determine how the initial frequency-independent initial strain value translates into a gravitational wave amplitude on horizon scales.  This is reflected in the 
behavior of the strain power spectrum as a function of momentum wave vector; in the inflationary case during the matter dominated era the spectrum goes as $3j_1(k\tau)/(k\tau)$ whereas in the radiation dominated era the dependence goes as $j_0$, (where $j_0$ and $j_1$ are spherical Bessel functions, $k$ is the comoving wavenumber, and $\tau$ is the conformal time).
These are equal magnitude as $k\tau \rightarrow 0$, but  differ by 15$\%$ at horizon crossing (with a larger amplitude during matter domination compared to radiation domination) (i.e. see \cite{krausswhite}).

In order to properly calculate the magnitude of waves inside the horizon one must take into account not just redshifting but also the possibility of additional sourcing. In  the case of self-ordering, waves are continually sourced on the horizon scale and below because fields begin ordering on this scale.  Indeed, this is what is responsible for the generation of horizon-sized modes in the first place   \cite{lmk1992, prl2008} .  This process continues for some period after horizon crossing, and given the differing dynamics during these periods there is no reason to expect that the results will be the same during these two epochs.


In the existing literature a number of  approximations are made for ease of calculation; these must be handled carefully in order to properly estimate this effect of the sourced radiation, and it is to this task we now turn. 
In \cite{prl2008} the evolution of the scale factor was smoothly interpolated between radiation dominated and matter dominated eras, and the evolution of modes were followed from their production (which was assumed to occur in the matter dominated era-- relevant to CMB polarization) to well inside the horizon. In later work, Fenu et al \cite{durrer} focused on waves entering the horizon during radiation domination. 

To investigate the effect of post-horizon crossing growth we plot the evolution of such a mode from its production at time $\tau_0$ to a time when its frequency is an order of magnitude smaller than that of the horizon, calculated by numerically solving the coupled scalar field evolution equations along with  the metric during radiation domination. As can be seen in Figure 1, not taking this subsequent evolution into account reduces the results by almost an order of magnitude. (The oscillatory behavior presumably reflects damped ringing of the wave modes upon sudden termination of the source generating them.)

\begin{figure}[ht]
\centering
\includegraphics[scale=.7]{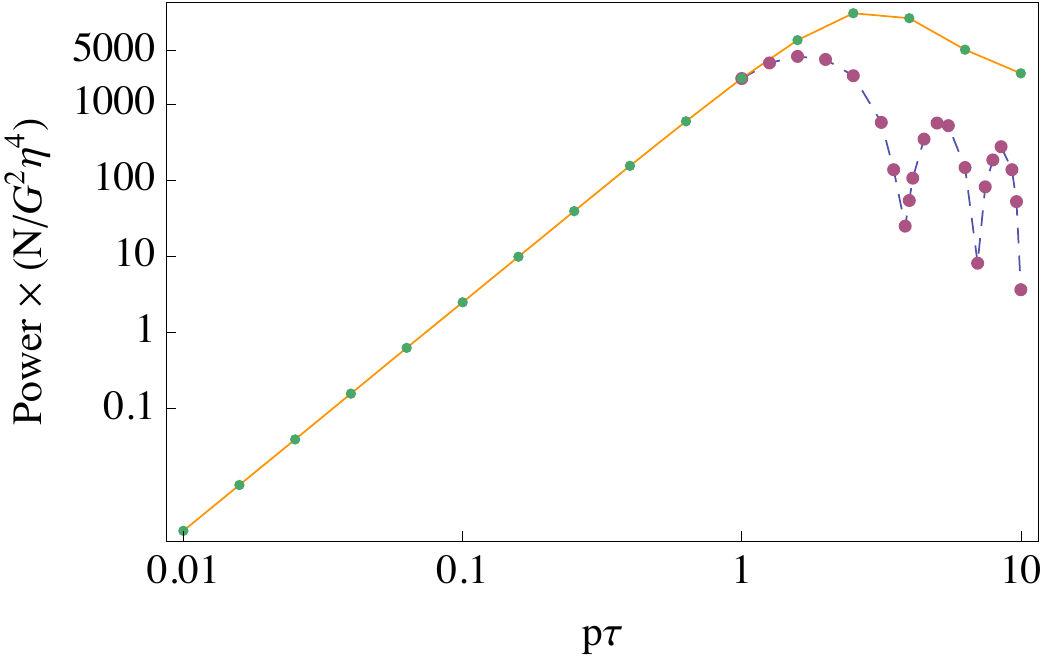}
\caption{\textit{Appropriately normalized power in Fourier modes with $p= x\tau$ as a function of $p\tau$.  The solid curve represents a calculation including full scalar field dynamics down to $p\tau =10$, while the dotted curve cuts of sourcing at horizon crossing, $p\tau=1$.}}
\label{crossing_plot}
\end{figure}

Using these detailed numerical estimates for the strain power generated during these two epochs we can compare the contribution to the energy density today in modes which crossed the horizon during radiation domination vs matter domination for backgrounds from self-ordering scalar fields and from Inflation, when these are normalized to give identical contributions in either the high frequency or low frequency domain today.  We assume, for the purposes of simple comparison, an exactly flat inflationary spectrum here.  

As can be seen in Figure 2, sourcing of modes inside the horizon is clearly more effective during matter domination than radiation domination, effectively combatting redshift damping for some time.  This may be due to the fact that during matter domination the energy momentum tensor of the relaxing scalar field differs more dramatically from that of the background matter density, therefore more effectively coupling the changing stress tensor to the production of gravitational waves.  In any case if direct detection can be accomplished with a precision of $\pm 50\%$, then the distinction between waves generated by cosmic phase transitions and those by inflation should be easily distinguishable.  In practice of course, the spectral tilt of the gravitational wave spectrum would have to be measured in the CMB in order to extrapolate expectations to frequencies appropriate for gravitational wave interferometers.   In this regard it is also important to note that if waves from cosmic phase transitions were incorrectly attributed to inflation one would infer an incorrect frequency dependent tilt dependence for the assumed inflationary modes.

\begin{figure}[ht]
\centering
\includegraphics[scale=.7]{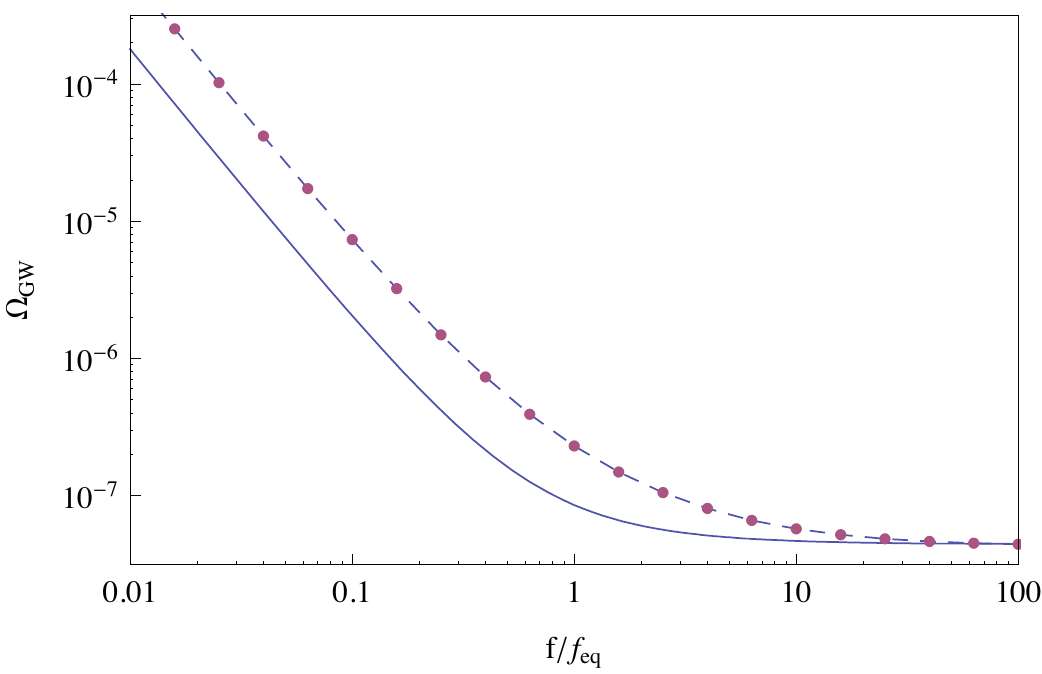}
\caption{\textit{The fractional energy density per logarithmic frequency interval as a function of frequency (normalized to the frequency of modes that cross the horizon at matter-radiation equality) for gravitational wave modes generated during inflation compared to that in modes generated by scalar field reordering, for transitions tuned to give the same gravitational wave amplitude for high frequency modes that enter the horizon during radiation domination. }}
\label{energy_plot}
\end{figure}

Inflationary plots such as figure 2 are generated by numerical solution of the source-free gravitational wave equation. Weinberg has derived a number of asymptotic analytic solutions to this equation
that are useful for checking numerical work and for semi-analytic estimates \cite{weinberg}. 
The formulae as derived apply
in the era\textcolor{blue}{s} of radiation and matter domination. Since the universe is now entering an epoch of vacuum domination
it is worthwhile to consider the extension of Weinberg's formulae past matter-vacuum equivalence. In the eras of
interest the relation between conformal time and scale factor is given by the Friedman equation
\begin{equation}
\tau = \frac{1}{H_n} \int_0^a d a' \; ( \Omega_R + \Omega_M a' + \Omega_{\Lambda} a'^4 )^{-1/2}
\label{eq:friedman}
\end{equation}
where $H_n$ is the present value of the Hubble constant, and $\Omega_i$ is the density parameter of the i'th form of energy. 
 It is convenient to define $a_{{\rm eq}} = (\Omega_R/\Omega_M)$ and $a_{\Lambda} = ( \Omega_M/\Omega_{\Lambda})^{1/3}$, the scale factors at radiation-matter
and matter-vacuum equivalence respectively. It is also helpful to define the wave-vectors $q_{{\rm eq}} = 
a_{{\rm eq}} H_{{\rm eq}}/\sqrt{2}$, $q_{\Lambda} = H_{\Lambda} a_{\Lambda}/ \sqrt{2}$ and $q_{\infty} = 
1/\tau_{\infty}$. It is easy to see
that the Friedman eq (\ref{eq:friedman}) simplifies to $\tau = 2 [ \sqrt{1 + y} - 1 ]/q_{{\rm eq}}$,with $\bf{y \equiv a/a_{eq}}$, in the eras of
radiation and matter domination ($a \ll a_{\Lambda}$) and to
$ \tau = \tau_{\infty} - 1/q_{\Lambda} z$ in the era of vacuum domination $a \gg a_{\Lambda}$. 

Weinberg showed that modes that enter the horizon in the era of radiation domination 
($q \gg q_{{\rm eq}}$) are given by the WKB approximate formula
\begin{equation}
D_q (\tau) = D_q (\tau_0) 
\frac{1}{\kappa y}
\sin q \tau.
\label{eq:radiationdomination}
\end{equation}
where $\kappa = q/q_{{\rm eq}}$ and $D_q (\tau_0)$ is the amplitude of the mode when it is far outside the horizon. 
The WKB formula applies once the mode enters the horizon ($ y \gg 1/\kappa$) and we find it applies not only
in the era of radiation and matter domination but it also 
continues to apply
through the present epoch deep into the era of vacuum domination until redshift $ z \ll Q $ where $Q = q/q_{\Lambda}$. 
We can continue the WKB solution eq (\ref{eq:radiationdomination}) 
into the indefinite future, $z \gg Q$, by connecting it to exact solutions to the gravitational wave equation that apply in 
the era of vacuum domination, $a \gg a_{\Lambda}$. We find
\begin{eqnarray}
D_q (\tau) = &
\nonumber \\
D_q (\tau_0) \frac{ q_{{\rm eq}} a_{{\rm eq}} }{ q_{\Lambda} a_{\Lambda} } \frac{1}{Q^2} 
\left[ \cos \frac{q}{q_{\infty}} D_2 (\frac{Q}{z}) - \sin \frac{q}{q_{\infty}} D_1 (\frac{Q}{z}) \right].
\label{eq:radiationfuture}
\end{eqnarray} 
This formula applies for all $z \gg 1$. Here $D_1 (x) = \sin x - x \cos x$ and $D_2 (x) = - x \sin x - \cos x$. 
The $D_2$ component
corresponds to modes that are pushed outside the horizon and freeze; $D_1$, to modes that continue
to decay forever. 
Eq (\ref{eq:radiationfuture}) shows that the eventual fate of the modes
is a sensitive oscillatory function of the wave-vector $q$ with period $q_{\infty}$. 

Similar formulae can be
written for modes that enter during the era of matter domination ($q_{\Lambda} \ll q \ll q_{{\rm eq}}$). The 
WKB solution in this case is
\begin{equation}
D_q (\tau) = - D_q (\tau_0) \frac{3}{4 \kappa^2 y} \cos q \tau
\label{eq:mattermodes}
\end{equation}
and gets extended to
\begin{eqnarray}
D_q (\tau) =  &&
\nonumber \\
D_q (\tau_0) \frac{3}{4} \frac{q_{{\rm eq}}^2 a_{{\rm eq}} }{ q_{\Lambda}^2 a_{\Lambda} } 
\frac{1}{Q^3} \left[ \cos \frac{q}{q_{\infty}} D_1 ( \frac{Q}{z} ) + \sin \frac{q}{q_{\infty} } D_2 (\frac{Q}{z} ) \right].
\label{eq:matterfuture}
\end{eqnarray} 

Finally, to compare CMB limits and direct detection experiments, one needs to take into account different normalization conventions.  CMB limits usually quote constraints on the strain directly, whereas traditionally constraints on direct signatures employ the energy density per logarithmic frequency interval. A consistent convention is to write the metric as $g^{\mu \nu} = a^2 ( \eta^{\mu \nu} + h^{\mu \nu}) $ where $\eta^{\mu \nu}$
is the Minkowski metric, $a$ is the scale factor, and $h$ denotes the strain perturbation. In this convention we define the Fourier transform of the strain by
\begin{eqnarray}
h_{ij}({\mathbf x},\tau) = \sum_{A = +,\times}{1 \over (2 \pi)^3} \int {d{\bf{k}}}\epsilon^{A}_{ij}({\bf{k}})h^{A} ({\mathbf k}, \tau)
(\tau)e^{i{\bf{kx}}}
\end{eqnarray}
where
\begin{eqnarray}
\sum_{i,j}\epsilon_{ij}^A\epsilon_{ij}^B = 2\delta^{AB},
\end{eqnarray}
and the strain power in the modes is defined via
\begin{eqnarray}
\delta^3({\mathbf k}-{\mathbf k'})\mathcal{P}(k, \tau)  \equiv & \frac{k^3}{2 \pi^2}
&  <h_+^{\ast} ({\bf k},\tau)h_+({\bf k}',\tau) 
\nonumber \\
& &
+ h_{\times}^{\ast}({\bf k},\tau)h_{\times}({\bf k}',\tau)>,
\label{eq:strainconvention}
\end{eqnarray}
where "+" and "$\times$" label the polarization states, 
and the gravitational energy density normalized to the critical energy density in the universe is defined via 
\begin{equation}
\Omega_{{\rm gw}} = \frac{1}{(2 \pi)^3} \int d \ln k \; \omega_{{\rm gw}} (k, \tau_{{\rm n}}) 
\label{eq:energydensityconvention}
\end{equation}
where $\omega_{{\rm gw}} (k)$ is the normalized energy density per log frequency interval and $\tau_{{\rm n}}$ is
the present conformal age of the Universe. From the
textbook formula relating the strain amplitude to the energy density of a gravitational plane wave it follows
\begin{equation}
\omega_{{\rm gw}} ( k ) = \frac{1}{6} \frac{k^2}{ H_n^2} P_h (p, \tau )
\label{eq:energystrainrelation}
\end{equation}
is the relationship between strain power and energy density. 

An alternative convention \cite{turner} is to define the Fourier transform with a factor of $(2 \pi)^{3/2}$ in the measure.
This convention produces strain Fourier amplitudes that are smaller by a factor of $(2 \pi)^{3/2}$ and a strain power that is smaller by a factor of $(2 \pi)^3$.  The
energy density per log frequency interval is still related to the strain power by eq (\ref{eq:energystrainrelation})
and it is therefore also smaller than the corresponding quantity in the first convention by a factor of $(2 \pi)^3$.
As a result, the relation between the total gravitational energy density and the energy density per log frequency interval
is then eq (\ref{eq:energydensityconvention}) with the factor of $(2 \pi)^3$ omitted. 

Ref \cite{prl2008} followed the first convention but with the twist that $h_+$ and $h_{\times}$ 
differ from the above definition by a factor of $-2$. As a result the
strain power and energy density calculated in that paper should be multiplied by a factor of $4$ to be consistent with the convention we define here.  
In that paper, however a comparison of the gravitational wave
energy density produced in this case with that due to inflation
at horizon crossing used a quoted result for the latter based not on the same convention used to derive results for cosmic phase transitions but rather that of  \cite{turner}.  As a result the statement there that the ratio of the energy densities in the two cases is larger than the naive dimensional estimate 
$ ( \eta / v N^{1/4} )^4$  by more than 
four orders of magnitude (where $\eta$ is the scale of the cosmic phase transition and 
$v^4$ is the energy density during inflation) should be modified by a factor of $ 4/ (2 \pi)^3 =(2 \pi)^{-3}$ (i.e. see \cite{durrer}). The correct statement is that the ratio of the gravitational energy density 
due to cosmic phase transitions to the inflationary energy density at frequencies accessible to 
direct detection is larger than the naive dimensional estimate by over two orders of magnitude.

The results here, which now allow a consistent and detailed comparison of strain power and associated energy densities a stochastic background of gravitational waves produced by cosmic phase transitions and inflation in two different regimes--for waves which entered the horizon during matter vs radiation domination--demonstrate the potential importance of comparing the magnitudes of the strain power in these two regimes.  As we have shown, if such a background is observed via a CMB polarization signal, a comparison with a possible direct gravitational wave detection at shorter wavelengths would provide a powerful tool in order to distinguish the origin of such a background, and thus provide an important diagnostic to investigate possible new GUT scale physics.

\end{document}